\documentclass[journal]{IEEEtran}
\usepackage{stmaryrd}
\usepackage{amsmath}
\usepackage{amsthm}
\usepackage{xfrac}
\usepackage{amssymb}
\usepackage{mathabx}
\usepackage{psfrag}
\usepackage{graphicx}
\usepackage{epstopdf}
\usepackage{todonotes}
\usepackage{cite}
\usepackage{algorithmic}
\usepackage{algorithm}
\usepackage{svg}
\usepackage{mathtools}
\usepackage{hyperref}
\usepackage{hyphenat}
\usepackage{booktabs}
\usepackage{accents}
\usepackage{bbm}
\usepackage{balance}
\usepackage{cleveref}

\newlength{\dhatheight}

\usepackage{float}
\usepackage{multirow}

\usepackage{xparse,mathtools}
\usepackage{subcaption} 

\newtheorem{lemma}{Lemma}

\newcommand\blfootnote[1]{%
  \begingroup
  \renewcommand\thefootnote{}\footnote{#1}%
  \addtocounter{footnote}{-1}%
  \endgroup
}

\begin{document}
%

\title{
Reducing Inter-user Interference: Precoding over
OFDM for Enhanced MTC}
\author{Karim A. Said, A. A. (Louis) Beex, Elizabeth Bentley, and Lingjia Liu
\thanks{K. Said, A. A. Beex and L. Liu are with Wireless@Virginia Tech, the Bradley Department of ECE at Virginia Tech, Blacksburg, VA. E. Bentley is with the Information Directorate of Air Force Research Laboratory, Rome NY. }
}

%




\maketitle
\begin{abstract}
In the physical layer (PHY) of modern cellular systems, information is transmitted as a sequence of resource blocks (RBs) across various domains with each resource block limited to a certain time and frequency duration. 
In the PHY of 4G/5G systems, data is transmitted in the unit of transport block (TB) across a fixed number of physical RBs based on resource allocation decisions. Using sharp band-limiting in the frequency domain can provide good separation between different resource allocations without wasting resources in guard bands. However, using sharp filters comes at the cost of elongating the overall system impulse response which can accentuate inter-symbol interference (ISI). In a multi-user setup, such as in Machine Type Communication (MTC), different users are allocated resources across time and frequency, and operate at different power levels. If strict band-limiting separation is used, high power user signals can leak in time into low power user allocations.
The ISI extent, i.e., the number of neighboring symbols that contribute to the interference, depends both on the channel delay spread and the spectral concentration properties of the signaling waveforms. 
 We hypothesize that using a precoder that effectively transforms an OFDM waveform basis into a basis comprised of prolate spheroidal sequences (DPSS) can minimize the ISI extent when strictly confined frequency allocations are used. Analytical expressions for upper bounds on ISI are derived. In addition, simulation results support our hypothesis.
 
\end{abstract}


\section{Introduction}

\blfootnote{Distribution A. Approved for public release: Distribution Unlimited: AFRL-2025-0072 on 07 Jan 2025 }
Orthogonal frequency division multiplexing (OFDM) has been selected as the physical layer waveform for the 5G NR standard \cite{lien20175g}, with the addition of multiple numerological variations to accommodate different application categories \cite{7794610}. One such category is machine-type-communication (MTC) or machine-to-machine (M2M) communications, which pertains to the notion of communication between devices uninitiated by humans \cite{mahmood2021machine}. MTC is expected to play a key role in a number of application domains such as public safety, health-care and industry automation, all of which involve reliance on a large number of wirelessly connected devices. 

In general, the typical traffic in M2M applications consists of low data rate transmissions comprised of short packet bursts. Certain applications impose additional requirements such as low latency and high reliability, namely mission critical M2M (mcM2M). The physical layer signaling waveform is a key factor in satisfying the specific application requirements in the presence of adverse channel effects. For example, in the presence of a massive number of M2M devices with diverse operation modes, inter-symbol interference (ISI) can result from waveforms with poor spectral confinement when sharply band-limiting filters are enforced.
One scenario is the interference from long range devices which transmit at high power resulting in high filter impulse response leakage in time. This leakage can cause significant ISI in high delay spread environments. An effect often referred to as leakage due to fractional delay taps, non-integer multiples of the sampling rate \cite{yilmaz2022control,7041655,7574382,sahin2013investigation}.

The 5G standard support for M2M comes in two forms: NBIoT and LTE-M, with bandwidth allocations of 180 kHz and 1.4 MHz, respectively \cite{8269112}. Even though 5G NR introduces the additional degree of freedom of controlling the numerology in the physical layer, more needs to be done to overcome the well known limitations of the OFDM waveform, such as the slow out-of-band spectral decay \cite{7023145}. 

 According to the Heisenberg uncertainty principle, high spectral tails can be tempered using pulse shapes with long support in the time-domain such as sinc and raised cosine pulse shapes \cite{harris1978use}. Many window shaping techniques exist in the literature to lower spectral lobes \cite{6521077} and improve performance in general \cite{rugini2006low}. Therefore, spectral confinement is achieved at the cost of time confinement. In other words, using pulse shaping to lower the spectral side-lobes will accentuate the effect of fractional delay spread. Furthermore, in the discrete setting, pulse shaping can lead to loss of orthogonality which will affect the conditioning of the equivalent channel matrix or even lead to singularity \cite{michailow2014generalized}. 
OFDM and other derivative waveforms such as filter bank multicarrier (FBMC) \cite{5753092} and orthogonal time frequency space (OTFS) \cite{7925924, 9508932} can be classified under the category known as Gabor frames \cite{sahin2013survey,strohmer2001approximation}. Gabor frames consist of a set of waveforms that are time and frequency shifted versions of a \textit{single} prototype pulse shape. Precoder based methods exist in the literature to modify the spectral properties of OFDM \cite{7490420, 9760053}. However, these have a high dependence on the data in the OFDM symbol which increases the complexity. In this work, we propose applying a fixed precoding to OFDM to obtain a new waveform basis set comprised of discrete prolate spheroidal sequences (DPSS). The resulting waveform consists of mutually orthogonal pulse shapes that are \textit{not} related by a time-frequency shift relationship, i.e, does not fall in the Gabor frame category. We show that the high concentration properties obtained from DPSS precoding can provide high immunity to leakage interference from adjacent signals in the time-domain. This is supported by our previous work \cite{10510883} on the optimal time confinement of DPSS in the presence of half-sample shifts. 

This work analyzes the inter-user interference due to fractional delay leakage in the time domain from high power users in M2M scenarios.  We show that due to the high concentration in both time and frequency of DPSS precoded OFDM, by only turning off a small percentage of \textit{edge waveform components} ($2\%$), ISI interference is reduced significantly compared to other waveforms.

The main contributions of this work are the following:
\begin{itemize}
    \item A mathematical framework for quantifying the effect of fractional time shifts on the energy spread for arbitrary waveforms. In this work we focus mainly on fractional time shift but the framework is applicable to fractional shifts in frequency as well.
    \item Upper bounds on inter-symbol interference for arbitrary signaling waveform. 
    \item A signaling waveform obtained by applying precoding to OFDM based on DPSS sequences and theoretical justification for its immunity to ISI resulting from fractional delay leakage.
\end{itemize}
 
 \section{System Model}

We adhere to a matrix framework for representing the discrete time input-output relations of operations at the transmitter and receiver, as well as channel effects.
Without loss of generality, at the transmitter a vector of information symbols $\mathbf{i} \in \mathbb{C}^{I \times 1}$ modulates OFDM sub-carriers represented by columns of the DFT matrix $\mathbf{F} \in \mathbb{C}^{N \times N}$,  to generate samples in the time domain represented by vector $\mathbf{x} \in \mathbb{C}^{N}$ where $\mathbf{x}=\mathbf{F}\mathbf{i}$. 
After undergoing the channel effects represented by time-varying impulse response matrix $\mathbf{H}\in \mathbb{C}^{N\times N}$ \cite{996901}, the signal vector $\mathbf{y}$ arrives at the receiver:
\begin{equation}\label{tx_mod_ch}
\mathbf{y}=\mathbf{H}\mathbf{x}+\mathbf{n}
\end{equation}
where $\mathbf{n}\in \mathbb{C}^{N}$ is an awgn noise vector.
A matched filtering operation (FFT for ODFM) is applied by correlating with the transmit waveform basis set:
\begin{equation}\label{eq_io}
\begin{split}
\mathbf{z}&=\mathbf{F}^H\mathbf{H}\mathbf{F}\mathbf{i}+\mathbf{F}^{H}\mathbf{n}\\
\end{split}
\end{equation}
Equation \eqref{eq_io} corresponds to the DFT modulation, channel effect, and receiver IDFT operations for a single OFDM symbol transmission. We introduce in \eqref{eq_io_precoding} the orthogonal precoding and un-precoding operations prior to the IFFT at the transmitter and post the IDFT at the receiver, respectively, by the matrix operation $\mathbf{S}$ and $\mathbf{S}^H$ where $\mathbf{S}\in \mathbb{C}^{N\times M}$ is an orthogonal matrix such that $N\leq M$. \begin{equation}\label{eq_io_precoding}
\begin{split}
\mathbf{z}&=\mathbf{O}^H\mathbf{H}\mathbf{O}\mathbf{i}+\mathbf{S}\mathbf{O}^{H}\mathbf{n}\\
\end{split}
\end{equation}
where $\mathbf{O}=\mathbf{FS}$.

We note that orthogonal precoding is equivalent to a basis transformation (incomplete basis for $M<N$) from OFDM to any other waveform basis, $\mathbf{O}$, such as single-carrier or even orthogonal time-frequency space (OTFS) \cite{8058662,zemen2018iterative}. In the case of OTFS, however, the transformation can best be described as a transformation basis consisting of a sequence of OFDM symbols to an OTFS frame. The correspondence between a sequence of $L$ OFDM symbols and OTFS with parameters $(L,N)$ is given by \eqref{OFDM_OTFS} \cite{8516353}
\begin{equation}\label{OFDM_OTFS}
\mathbf{I}_L\otimes \mathbf{F} \text{ (OFDM)}, \quad \mathbf{F}\otimes \mathbf{I}_L \text{ (OTFS)}
\end{equation}
Extending \eqref{eq_io_precoding} to the case of a stream of $L$ OFDM symbols,  
\begin{equation}\label{eq_io_precoding_stream}
\begin{split}
\bar{\mathbf{z}}&=(\mathbf{I}_L\otimes\mathbf{O})^H\mathbf{H}(\mathbf{I}_L\otimes\mathbf{O})\bar{\mathbf{i}}+(\mathbf{I}_L\otimes\mathbf{O})^H\bar{\mathbf{n}}\\
\end{split}
\end{equation}
where $\bar{\mathbf{z}}=[\mathbf{z}_1^T,..,\mathbf{z}_L]^T$, $\bar{\mathbf{i}}=[\mathbf{i}_1^T,..,\mathbf{i}_L]^T$, $\bar{\mathbf{n}}=[\mathbf{n}_1^T,..,\mathbf{n}_L]^T$, $\mathbf{I}_L$ is an identity matrix of size $L\times L$ and $\otimes$ is the Kronecker product.

\begin{figure}
\centering 
\includegraphics[width=1\linewidth]{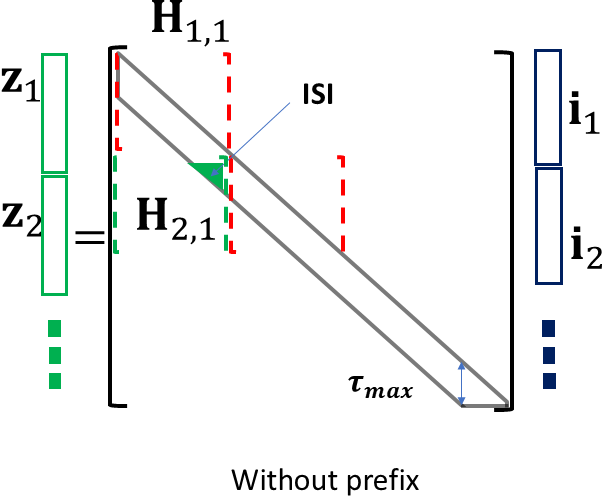}
\caption{Time-varying impulse response (TV-IR) channel matrix with integer delay taps depicted by diagonal parallelogram. Input blocks with no prefix shown on the right and corresponding output block on the left.} \label{ch_io_vis}
\end{figure}

\begin{figure}
\centering 
\includegraphics[width=1\linewidth]{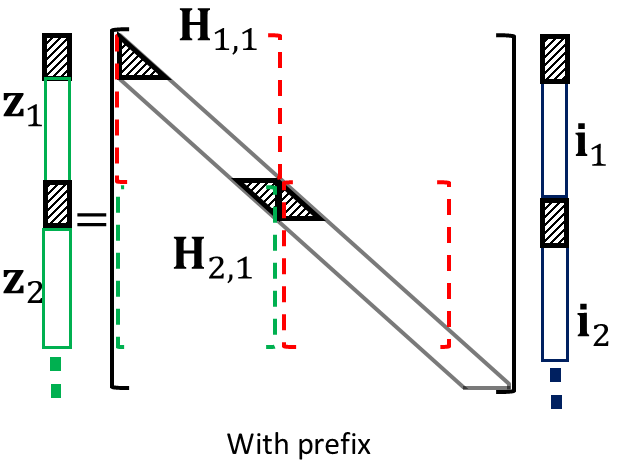}
\caption{Time-varying impulse response (TV-IR) channel matrix with integer delay taps depicted by diagonal parallelogram. Input blocks with prefix shown on the right and corresponding output block on the left after prefix is removed. } \label{ch_io_vis_cp}
\end{figure}
The input-output relationship given in \eqref{eq_io_precoding_stream}  is visualized in Fig.\ref{ch_io_vis}. The impulse response channel matrix $\mathbf{H}$ is strictly banded and the region of non-zero values is depicted  by the thin and tall parallelogram having a small side length $\tau_{max}$, the delay spread of the channel. The input consisting of the sequence of symbols $\mathbf{i}_1, \mathbf{i}_2,.. $  depicted by blue rectangles on the right is multiplied by the matrix to generate the sequence of output vectors $\mathbf{z}_1,\mathbf{z}_2,..$, depicted by green rectangles on the left. The green triangle depicts the part of the channel matrix that causes ISI from input symbol $\mathbf{i}_1$ into  output vector $\mathbf{z}_2$.

On a symbol by symbol level, \eqref{eq_io_precoding_stream} can be broken down into the following:
 \begin{equation}\label{eq_io_one_symbysym}
\begin{split}
\mathbf{z}_l=\mathbf{O}^H\mathbf{H}_{l,l}\mathbf{O}\mathbf{i}_l+\sum_{l'\neq l}\mathbf{O}^H\mathbf{H}_{l,l'}\mathbf{O}\mathbf{i}_{l'}+\mathbf{O}^H\mathbf{n}_l, \\\quad l,l'=0,..,L-1
\end{split}
\end{equation}
where $\left[\mathbf{H}_{l,l'}\right]_{n,n'}=\left[\mathbf{H}\right]_{lN+n,l'N+n'}, n,n'=0,..,N-1$.

The second sum in \eqref{eq_io_one_symbysym} represents the inter-symbol interference. Knowing the length of the delay spread $\tau_{max}$, prepending the input symbols with a prefix (zero prefix or cyclic prefix) of length $\tau_{max}$ can eliminate the ISI. In Fig. \ref{ch_io_vis_cp}, the channel output corresponding to the black shaded triangles, denoted by black shaded rectangles, are ignored at the receiver. This effectively removes the black shaded triangles from the channel matrix. Now, the square bracketed sub-matrices $\mathbf{H}_{l,l'}, l\neq l'$ are canceled out. To incorporate the effect of adding a prefix to the input symbols, and removing the output corresponding to the prefix at the channel output, \eqref{eq_io_precoding_stream} is modified as follows:
\begin{equation}\label{eq_io_precoding_stream_prfx}
\begin{split}
\bar{\mathbf{z}}&=(\mathbf{I}_L\otimes\mathbf{O}_r)^H\mathbf{H}(\mathbf{I}_L\otimes\mathbf{O}_t)\bar{\mathbf{i}}+(\mathbf{I}_L\otimes\mathbf{O}_r)^H\bar{\mathbf{n}}\\
\end{split}
\end{equation}
where $\mathbf{O}_r=[\mathbf{G}_z^T,\mathbf{O}^T]^T$, $\mathbf{O}_t=[\mathbf{G}_c^T,\mathbf{O}^T]^T$ or $\mathbf{O}_t=[\mathbf{G}_z^T,\mathbf{O}^T]^T$ , $\mathbf{G}_z=\mathbf{0}_{\tau_{max}\times N}$ and $\mathbf{G}_c = \mathbf{O}[N-\tau_{max}+1:N,:]$ (rows in the range $N-
\tau_{max}+1:N$), and the symbol by symbol level relationship becomes
 \begin{equation}\label{eq_io_one_symbysym_prfx}
\begin{split}
\mathbf{z}_l=\mathbf{O}_r^H\mathbf{H}_{l,l}\mathbf{O}_t\mathbf{i}_l+\mathbf{O}_r^H\mathbf{n}_l, \\\quad l'=0,..,L-1
\end{split}
\end{equation}
where the term  $\sum_{l'\neq l}\mathbf{O}_r^H\mathbf{H}_{l,l'}\mathbf{O}_t\mathbf{i}_{l'}$ is not included due the effect of the prefix, which causes it to be zero. 

However, zero ISI due to the prefix rests on the assumption that the impulse response matrix $\mathbf{H}$ is strictly banded. This requires the channel delay taps  $\tau_i\in [0,\tau_{max}]$ to be integer multiples of the sampling period of the input signal, i.e., $W\tau_i \in \mathbb{Z}$, where $W$ is the signal bandwidth in units of Hz, and $\tau_i$ is the delay of the $i$-th tap in seconds. In practice, this is rarely the case and the normalized delay of the taps is fractional. Thus, a channel with $P$ delay taps at non-integer delays, and amplitudes $g_p \in \mathbb{C}$, when acting on a strictly band-limited signal is equivalent to an infinite number of \textit{integer} delay taps as follows \cite{482137}:
\begin{equation}\label{eq_intgr_ch}
    h[n]=\sum_{p=0}^{P-1}g_p \text{sinc}(n-W\tau_p), n\in \mathbb{Z}
\end{equation}
where $W$ is the band-width of the signal.
As a result, the channel matrix $\mathbf{H}$ will no longer be a strictly banded matrix, but a full matrix with non-zero elements on $N-1$ super and $N-1$ sub-diagonals. As a result, there will be residual ISI in spite of the usage of a prefix. The level of the residual ISI will depend on the time extent of the impulse response of the filter used to limit the bandwidth of the input signal, i.e., the more confined the bandwidth, the higher the ISI level.

\section{Impact of precoding on ISI }
In the previous section it was demonstrated that a symbol prefix does not guarantee the elimination of ISI in the presence of fractional delay taps. Knowing that fractional delay tap leakage is a manifestation of the poor time-frequency confinement of the signaling waveform, in this section we analyze the impact of waveform transformation by different types of precoding on leakage due to fractional delay.

For a channel comprised of $P$ discrete specular paths, the effect of each path can decomposed into two factors \cite{1608558,hlawatsch2011wireless}, a modulation factor (Doppler shift) and a convolution factor (time shift). The first factor is a diagonal matrix denoted by $\mathbf{D}(\nu_p)$, and the second factor is a Toeplitz matrix denoted by $\mathbf{T}(\tau_p)$
\begin{equation}\label{ch_matrix}
\begin{split}
\mathbf{H}&=\sum_{p=0}^{P-1}g_p\mathbf{D}(\nu_p)\mathbf{T}(\tau_p)
\end{split}
\end{equation}
where $\left[\mathbf{D}(\nu_p)\right]_{l,k}=e^{j2\pi l\nu_p}\delta[l-k]; l,k=0,..,N-1$;  $\nu_p$ is the normalized Doppler frequency of the $p$-th path, $[\mathbf{T}(\tau_p
)]_{l,k}=\frac{\sin\pi(l-k-\tau_p)}{\pi(l-k-\tau_p)}, l,k=0,..,N-1$ is the delay effect for normalized delay $\tau_p$ for the $p$-th path, and $g_p$ is the path gain.

Substituting in \eqref{eq_io_one_symbysym},
 \begin{equation}\label{eq_io_one_symbysym_beta}
\begin{split}
\mathbf{z}_l=\sum_{p=0}^{P-1}g_p\mathbf{O}^H\mathbf{H}_{l,l}\mathbf{O}\mathbf{i}_l+\sum_{l'\neq l}\boldsymbol \beta_{l,l'}\mathbf{i}_{l'}+\mathbf{O}^H\mathbf{n}_l, \\\quad l =0,..,L-1
\end{split}
\end{equation}
where $\boldsymbol \beta_{l,l'}$ is the ISI transfer function from input block $l'$ to output block $l$
\begin{equation}\label{ISI_Beta}
\begin{split}
\boldsymbol \beta_{l,l'}&=\sum_{p=0}^{P-1}g_p\mathbf{O}^H\mathbf{H}_{l,l'}\mathbf{O}\\
&=\sum_{p=0}^{P-1}g_p\mathbf{O}^H\mathbf{D}_{l}(\nu_p)\mathbf{T}_{l-l'}(\tau_p)\mathbf{O}
\end{split}
\end{equation}
where $[\mathbf{D}_{l}(\nu_p)]_{n,n}=[\mathbf{D}(\nu_p)]_{lN+n,lN+n}, n=0,..,N-1$, $[\mathbf{T}_{l}(\nu_p)]_{n,n'}=[\mathbf{T}(\nu_p)]_{lN+n,n'}, n,n'=0,..,N-1$.

Splitting the channel into delay and Doppler factors, we can see that the \textit{extent} of ISI depends mainly on the delay factor  $\mathbf{T}_{l-l'}(\tau_p)$. Thus, in what follows we focus our analysis on purely delay dispersive channels, i.e., $\nu_p=0$, or quasi static scenarios where channel variation over one symbol is negligible. In this case, $\boldsymbol \beta_{l,l'}$ simplifies  to
 \begin{equation}\label{ISI_Beta_pure_dly}
\begin{split}
\boldsymbol \beta_{l,l'}&=\sum_{p=0}^{P-1}g_p\mathbf{O}^H\mathbf{T}_{l-l'}(\tau_p)\mathbf{O}\\
\end{split}
\end{equation}


We analyze the ISI matrix $\boldsymbol \beta_{l,l'}$ element-wise by substituting with the definition of $\mathbf{T}_{l-l'}(\tau_p
)$:
\begin{equation}\label{delay_factor_matrix_element}
\begin{split}
&[\boldsymbol \beta_{l,l'}]_{r,s}\\
&=\sum_{p=0}^{P-1}\sum_{n,m=0}^{N-1} g_p \mathbf{o}_r^*[n] \mathbf{o}_s[m]\text{sinc}(n-m-(l'-l)N-\tau_p)\\
&=\sum_{p=0}^{P-1}\sum_{q=-N+1}^{N-1} g_p\mathbf{C}_{rs}[q]\text{sinc}\left(q-(\tau_p+(l'-l)N)\right)\\
\end{split}
\end{equation}
where $\mathbf{C}\in \mathbb{C}^{M\times M\times (2N-1)}$ is the cross-correlation tensor whose $r,s$ entry is the sequence $\mathbf{C}_{rs}[q]=\sum_{n=\max(-(N-1)/2+q,-(N-1)/2)}^{\min((N-1)/2+q,(N-1)/2)}\mathbf{o}_r^*[n]\mathbf{o}_s[n-q]$, where $\mathbf{o}_r$ is the $r$-th column of $\mathbf{O}$, and $\text{sinc}(x)=\frac{\sin \pi x}{\pi x}$. 

We note that 
the inner sum in the second line of \eqref{delay_factor_matrix_element} is the $r-s$-th cross-correlation sequence, $\mathbf{C}_{rs}$, shifted by $\tau_p+(l'-l)N$. $\mathbf{C}_{rs}[q]$ is an index limited sequence which when shifted by a fractional value $\tau_p$ results in a sequence that is infinite in extent. This elongation effect is the underlying cause of ISI that extends past the cyclic/zero prefix. To the best of our knowledge, no works exist to simplify the last line in \eqref{delay_factor_matrix_element} to an analytical closed form expression for fractional shifts $\tau_p$. Finding such an analytical expression would enable us to quantify the ISI energy which can provide us some measure of the expected degradation in SER performance. 

In what follows we pursue analytical expressions for upper bounding ISI energy for arbitrary waveforms.
We set up the mathematical framework that will lead to our final result in the following section.
\subsection{Leakage due to fractional delay taps}  
The inner sum in  \eqref{delay_factor_matrix_element} is essentially a composition of a bandlimit then time shift operation.  We define the operator notation $\lbrace \mathcal{B}_W^{\tau} r\rbrace[n]$ in \eqref{delay_factor_matrix_XCORR}, which represents the operation on a sequence $r[n]$ that outputs a sequence limited in frequency to half-bandwidth $W$ (normalized), scaled by $1/W$, then shifted by $\tau\in \mathbb{R}$.  We rewrite \eqref{delay_factor_matrix_element} after substituting with $\lbrace \mathcal{B}_W^{\tau} r\rbrace[n]$

\begin{equation}\label{delay_factor_matrix_element_2}
\begin{split}
[\boldsymbol \beta_{l,l'}]_{r,s}
&=\sum_{p=0}^{P-1} g_p\left\lbrace\mathcal{B}_{0.5}^{\tau_p} \mathbf{C}_{rs}\right\rbrace[(l'-l)N]\\
\end{split}
\end{equation}
where
\begin{equation}\label{delay_factor_matrix_XCORR}
\begin{split}
\left\lbrace\mathcal{B}_W^{\tau} \mathbf{C}_{rs}\right\rbrace &\triangleq \sum_{q=-N+1}^{N-1} \mathbf{C}_{rs}[q]\text{sinc}\left( Wq-\tau\right)\\
\end{split}
\end{equation}

Without loss of generality, the average ISI energy affecting a given symbol can be found by averaging the ISI component, second term in \eqref{eq_io_one_symbysym_beta}, at $l=0$:
\begin{equation}\label{ISI_energy}
\begin{split}
\text{E}^{ISI}&=\mathbb{E}\left\lbrace\left|\sum_{l'\neq 0}\boldsymbol \beta_{0,l'}\mathbf{i}_{l'}\right|^2\right\rbrace\\
&=\sum_{l',l\neq 0, }\mathbb{E}\left\lbrace\mathbf{i}_{l'}^H\boldsymbol\beta_{0,l'}^H\boldsymbol\beta_{0,l}\mathbf{i}_{l}\right\rbrace\\
&=\sum_{l'\neq 0}\text{Tr}\left(\mathbb{E}\left(\boldsymbol\beta_{0,l'}^H\boldsymbol\beta_{0,l'}\right)\right)\\
\end{split}
\end{equation}
where $\mathbb{E}\left(\mathbf{i}_{l}\mathbf{i}_{l'}^H\right)=\mathbf{I}_M\delta(l-l')$.
Substituting into \eqref{ISI_energy} using \eqref{delay_factor_matrix_element_2}
\begin{equation}\label{ISI_energy2}
\begin{split}
&\text{E}^{ISI}\\
&=\sum_{l'\neq 0}\text{Tr}\left(\mathbb{E}\left(\boldsymbol\beta_{0,l'}^H\boldsymbol\beta_{0,l'}\right)\right)\\
&=\sum_{p,q=0}^{P-1}\sum_{l'\neq 0}\sum_{r,s}\mathbb{E}\left(   g_pg_q^*\left\lbrace\mathcal{B}_{0.5}^{\tau_p} \mathbf{C}_{rs}\right\rbrace[l'N]\left\lbrace\mathcal{B}_{0.5}^{\tau_q} \mathbf{C}_{rs}\right\rbrace^*[l'N]  \right)\\
&=\sum_{p=0}^{P-1} \sigma_p^2\sum_{r,s} \sum_{l'\neq 0} \left|\left\lbrace\mathcal{B}_{0.5}^{\tau_p} \mathbf{C}_{rs}\right\rbrace[l'N]  \right|^2\\
&=\sum_{p=0}^{P-1} \sigma_p^2\text{E}^{ISI}(\tau_p)\\
\end{split}
\end{equation}
where $\mathbb{E}(g_pg_q)=\sigma_p^2\delta(p-q)$, and $\text{E}^{ISI}(\tau_p)$ is the per-path ISI energy of the $p$-th path. 
 We note that in the presence of a zero guard prefix of length $g>\tau_p$, $\text{E}^{ISI}(\tau_p)=0$ for $\tau_p \in \mathbb{Z} $.

The waveform dependence of per-path ISI energy is given by the term $\sum_{l'\neq 0} \left|\left\lbrace\mathcal{B}_{0.5}^{\tau_p} \mathbf{C}_{rs}\right\rbrace[l'N]  \right|^2 $ in the fourth line of \eqref{ISI_energy2}.

We provide a definition followed by an upper bound which we refer to as the \textit{band-limited correlation tail energy bound} $\text{E}_{BCT}$.

\newtheorem{definition}{Definition}
\begin{definition}[Tail Energy of a Sequence]\label{def1}
  $\bar{E}_{l}$ for $l\in \mathbb{Z}^{+}$:
 \begin{equation}\label{dbl_side_tail_E}
\bar{E}_{l}(r) \triangleq \sum_{n=-\infty}^{-l-1}|r^2[n]|+\sum_{n=l+1}^{\infty}|r^2[n]|
\end{equation}
\end{definition}

\begin{equation}\label{ISI_wvfrm_dpndnc}
\begin{split}
  \sum_{l'\neq 0} \left|\left\lbrace\mathcal{B}_{0.5}^{\tau_p} \mathbf{C}_{rs}\right\rbrace[l'N]  \right|^2&\leq\bar{E}_{(N-1)}(\mathcal{B}_{0.5}^{\tau_p} \mathbf{C}_{rs})\\
  &\leq\bar{E}_{N-1-\lfloor\tau_p\rfloor}(\mathcal{B}_{0.5}^{\Delta\tau_p} \mathbf{C}_{rs})\\
  &\leq\bar{E}_{N-1-\lfloor\tau_p\rfloor}(\mathcal{B}_{0.5}^{0.5} \mathbf{C}_{rs})\\
  &\leq \text{E}_{BCT}(r,s)\\
\end{split}
\end{equation}
where $ \Delta \tau_p=\tau_p-\lfloor\tau_p\rfloor$, and
\begin{equation}\label{E_BCT}
\text{E}_{BCT}(r,s)\triangleq\bar{E}_{N-1}(\mathcal{B}_{0.5}^{0.5} \mathbf{C}_{rs}) 
\end{equation}
The transition from the second to third line in \eqref{ISI_wvfrm_dpndnc} is based on our conjecture that a half sample shift results in the highest tail energy compared to all other fractional sample shifts $\tau_p<0.5$.

We note that $\text{E}_{BCT}(r,s)$ depends only on the precoding and has no dependence on channel parameters. It is used in the following section, as a metric to compare between different precoding options. However, we note that the numerical evaluation of $\text{E}_{BCT}(r,s)$ involves a sum with infinite limits. The following tight bound, which can be found in \cite{10510883} involves a finite number of computations and can be used as a proxy.
\begin{equation}\label{BCTE_bound}
\begin{split}
  \text{E}_{BCT} \leq 4\sum_{l=0}^{4N} \left|c_{r,s}(l;N)\right|^2\lambda_l(1-\lambda_l)\\
\end{split}
\end{equation}
where $ c_{r,s}(l;N)=\sum_{n=-2N_g}^{2N}\mathbf{C}_{rs}[n]s_l^{(0.25,4N+1)}[2n]$, where $s_l^{(0.25,4N+1)}$ is the $l$-th member sequence from the DPSS set with parameters $0.25,4N+1$, and $\lambda_l$ is its corresponding eigenvalue.

\subsection{Impact of Precoding on Leakage Energy: Waveform Basis Subset Selection}

\begin{figure}
\centering 
\includegraphics[width=1\linewidth]{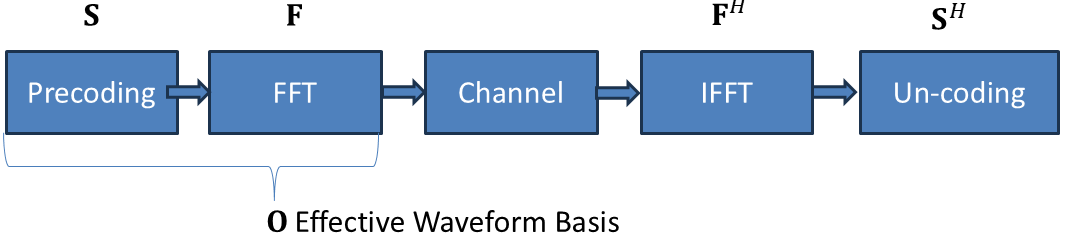}
\caption{Precoding and the resulting effective waveform basis.}\label{precoding}
\end{figure}

In what follows, we evaluate the impact of precoder choice on the per-path ISI using the obtained bound \eqref{E_BCT}. Each precoder results in a waveform basis and a corresponding cross-correlation tensor. Three precoding choices are considered: No precoding (OFDM waveform), DFT precoding (SC-FDMA), and DPSS precoding that results in a waveform consisting of discrete prolate spheroidal sequences as its basis components. Fig. \ref{precoding} illustrates the precoding operation at the transmitter, represented by matrix $\mathbf{S} \in \mathbb{C}^{N \times M}$ and the resulting effective waveform.

\hfill
\subsubsection{No precoding-OFDM }

The precoding matrix used is $\mathbf{S} = [\mathbf{0}_{(N-M)/2}^T,\mathbf{I}_M^T,\mathbf{0}_{(N-M)/2}^T]^T$ and the effective waveform basis is $\mathbf{O}=[\mathbf{o}_{0},..,\mathbf{o}_{M-1}]$, where $\mathbf{o}_m[n]=\frac{1}{\sqrt{N}}e^{\frac{j2\pi (m-\frac{N+1}{2})n}{N}} ; \quad m=0,..,M-1 ; \quad n=-(N-1)/2,..,(N-1)/2$. For $M<N$, the effective waveform basis consists of all subcarriers except for the edge subcarriers which are turned off.  Nulling edge subcarriers is a common practice to reduce out-of-band emissions (OOB) \cite{huang2015out}.

We start by finding the elements of the Cross-correlation tensor denoted by $\mathbf{C}^{F}_{r,s}[q]$. Knowing that $\mathbf{C}^{F}_{r,s}[q]=\mathbf{C}^{*F}_{s,r}[-q]$, we focus only on the cases $q\geq 0 \quad \forall r,s$
\begin{equation}\label{crss_ambg_FD}
\begin{split}
 \mathbf{C}^{F}_{r,s}[q] &= \frac{1}{N} \sum_{n=p-(N-1)/2}^{(N-1)/2}e^{-j\frac{2\pi r n}{N}} e^{j\frac{2\pi s (n-p)}{N}} \quad \text{for } p\geq 0\\
 &= \frac{1}{N} e^{-j\frac{2\pi (r+s) p/2}{N}}\sum_{n=-(N-1-p)/2}^{(N-1-p)/2}e^{-j\frac{2\pi (r-s) n}{N}} \\
 &= \frac{1}{N} e^{-j\frac{2\pi (r+s) p/2}{N}}\frac{\sin(\frac{N-p}{N+1}\pi (r-s))}{\sin(\frac{\pi(r-s)}{N})}\\
 \end{split}
\end{equation}

\begin{figure}
\label{F_sig_XCORR}
\centering
\begin{subfigure}[b]{\linewidth}
\includegraphics[width=\linewidth]{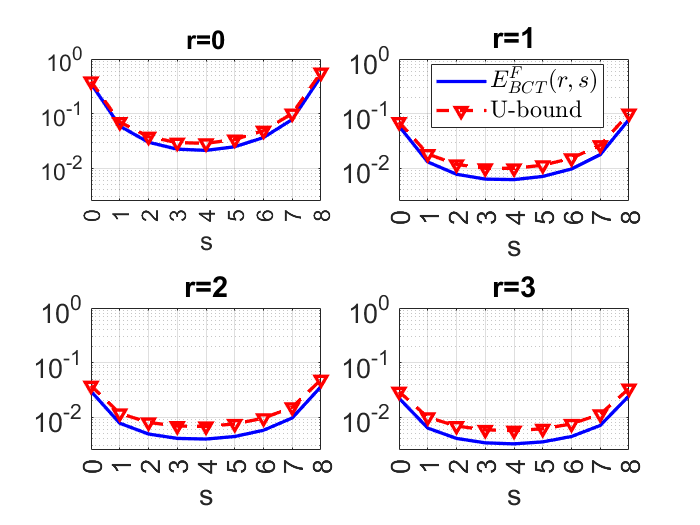}
\end{subfigure}
\begin{subfigure}[b]{\linewidth}
\includegraphics[width=\linewidth]{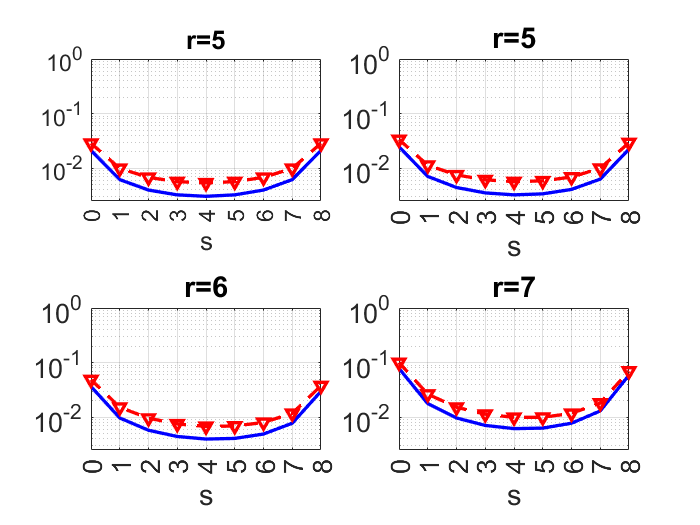}
\end{subfigure}
\begin{subfigure}[b]{0.5\linewidth}
\includegraphics[width=\linewidth]{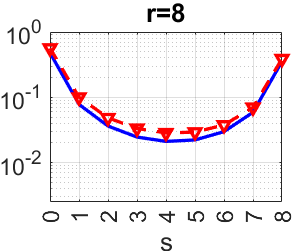}
\end{subfigure}
\caption{Plot of $\text{E}_{BCT}^F(r,s)$ in blue for the OFDM waveform, i.e, no precoding. Upper bound in red.}\label{F_sig_XCORR_F}
\end{figure}

For $N=9$, we numerically compute $E_{BCT}^F(r,s)$ by substituting using \eqref{crss_ambg_FD} in \eqref{E_BCT} and plot the computed values across $s$ on the x-axis in the $r-$th sub-figure of Fig. \ref{F_sig_XCORR_F}.  using the blue curve. In all sub-figures, the two highest values happen at $s=0,8$; furthermore, across all sub-figures, the curves in sub-figures $r=0, 8$ uniformly exceed the curves in other sub-figures. This can be explained by the fact that sub-waveforms $0, 8$ correspond to the highest two frequencies $\pm\frac{4\pi}{9}$, i.e., closest to the edge of the band-limiting filter, and thus most affected by it.

\hfill
\subsubsection{DFT precoding - SC-FDMA}
The second type of precoding we refer to as DFT precoding, which is similar to what is used in SC-FDMA.  The precoding matrix is given by $\mathbf{S} = [\mathbf{0}_{(N-M)/2}^T,\mathbf{F}_M^T,\mathbf{0}_{(N-M)/2}^T]^T$, where $\mathbf{F}_M$ denotes a DFT matrix of size $M$.
The $m$-th column of the effective waveform basis, $\mathbf{O}$, is given by \eqref{sig_TD_diric},
\begin{equation}\label{sig_TD_diric}
\begin{split}
\mathbf{o}_m[n]&=\frac{1}{\sqrt{\eta}N}\sum_{m'=-(M-1)/2}^{(M-1)/2} e^{j2\pi\frac{nm'}{N}}e^{-j2\pi\frac{m'm}{M}} \\
&=\frac{1}{\sqrt{\eta}N}\sum_{m'= -(M-1)/2}^{(M-1)/2} e^{j2\pi\frac{\left(n-\frac{m}{\eta}\right)m'}{N}}\\
&=\sqrt{\eta}\frac{\sin\left(M \frac{\pi}{N} \left(n-\frac{m}{\eta}\right)\right)}{M\sin\left(\frac{\pi}{N}\left(n-\frac{m}{\eta}\right)\right)}
\end{split}
\end{equation}
where $\eta=\frac{M}{N}$ controls the percentage of active waveforms which are fully confined within the frequency band $[-0.5\eta,0.5\eta]$. Note that for $\eta=1$, $\mathbf{o}_m[n]=\delta(n-m)$.

\begin{equation}\label{crss_ambg_TD}
\begin{split}
 &\mathbf{C}^T_{r,s}[p] \\
 &= \frac{1}{\eta N^2} \times \\
 &\sum_{l',k'=-\frac{L-1}{2}}^{\frac{L-1}{2}}e^{j\frac{2\pi rl'}{L}}\frac{\sin(\frac{N-p}{N}\pi (l'-k'))}{\sin(\frac{\pi(l'-k')}{N})}e^{-j\frac{\pi (l'+k') p}{N}}e^{j\frac{-2\pi k'(s+p\eta)}{L}}  \\
  \end{split}
\end{equation}
See Appendix II.
\begin{figure}
\label{T_sig_XCORR}
\centering
\begin{subfigure}[b]{\linewidth}
\includegraphics[width=\linewidth]{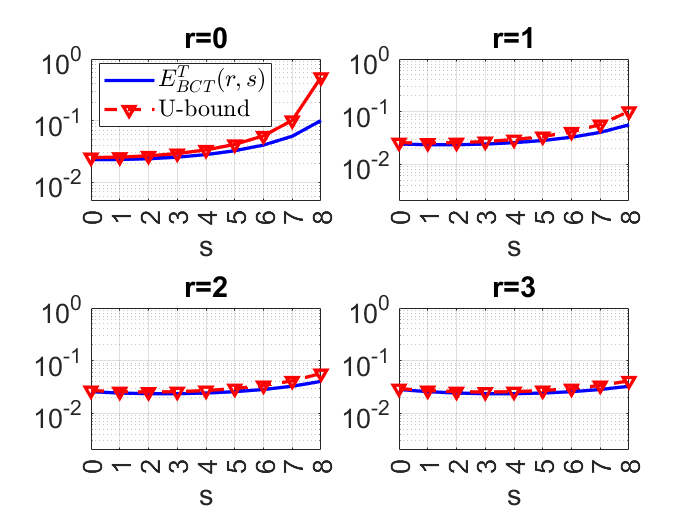}
\end{subfigure}
\begin{subfigure}[b]{\linewidth}
\includegraphics[width=\linewidth]{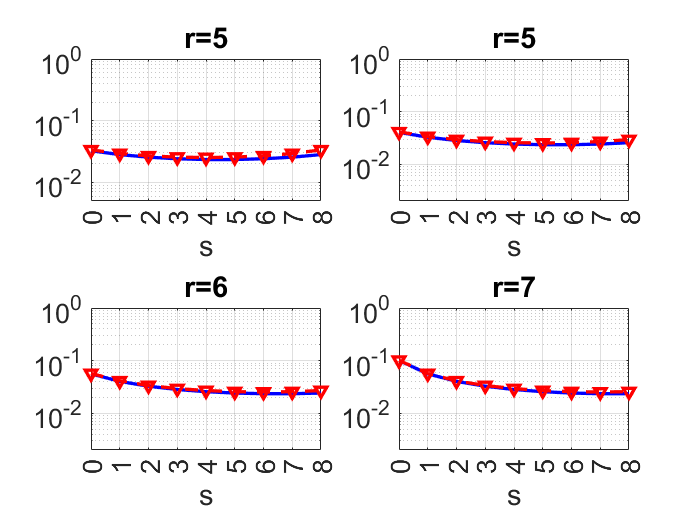}
\end{subfigure}
\begin{subfigure}[b]{0.5\linewidth}
\includegraphics[width=\linewidth]{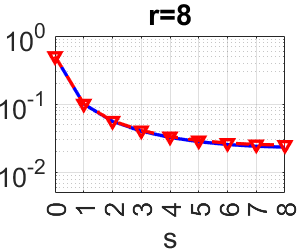}
\end{subfigure}
\caption{Plot of $\text{E}_{BCT}^T(r,s)$ in blue for the SC-FDM waveform, i.e, DFT precoding. Upper bound in red.}\label{F_sig_XCORR_I}
\end{figure}

$\text{E}_{BCT}^T(r,s)$ is plotted across $s$ on the x-axis in the $r-$th sub-figure of Fig. \ref{F_sig_XCORR_I}.  using the blue curve and the upper bound is plotted in red. Curves in sub-figures $r=0, 8$ uniformly exceed the curves in the other sub-figures with the highest two points in each sub-figure being at $s=0, 8$. Time domain sub-waveforms are versions of a proto-type pulse shape that are shifted in time, $s=0$ being left-most and $s=8$ being right-most in time. The points $r,s=0,8$ are cross-correlations involving sequences at the edges of the time index set $-(N-1)/2,..,(N-1)/2$.

\hfill
\subsubsection{DPSS precoding - DPSS Waveform Basis}
Discrete prolate spheroidal sequences are solutions of eigenvalue equation \eqref{DPSS_eig_eqn}\cite{slepian1978prolate}: 
\begin{equation}\label{DPSS_eig_eqn}
\lambda_l\mathbf{p}_l^T[n]=\sum_{m=-N/2}^{N/2}\mathbf{p}_l^S[m]\frac{\sin 2\pi W(n-m)}{\pi(n-m)}
\end{equation}
The precoding matrix used is comprised of the DFT of DPSSs, i.e.,  $\mathbf{S}=\mathbf{F}^H\mathbf{P}$. Such DFTs are commonly known in the literature as discrete prolate spheroidal wave functions (DPSWFs) \cite{slepian1978prolate,said2023non}.
The chosen set of DPSSs $\mathbf{P}=[\mathbf{p}_0,..,\mathbf{p}_{M-1}]$ where $\mathbf{p}_m$
are solutions of \eqref{DPSS_eig_eqn} for $W=0.5^-$,where $0.5^{-}$ denotes a value infinitesimally close to $0.5$ from below.  Thus, the effective waveform basis consists of discrete prolate spheroidal sequences, i.e., $\mathbf{O}=\mathbf{F}\mathbf{S}=\mathbf{P}$.

DPSSs are known to be bandlimited sequences with optimal time concentration. In our recent work it has been proven that the DPSS time-concentration property transfers to a correlation-concentration \cite{9647920}. Nevertheless, those correlation concentration results did not involve fractional shifts, unlike those being used in this work.

\begin{figure}
\label{S_sig_XCORR}
\centering
\begin{subfigure}[b]{\linewidth}
\includegraphics[width=\linewidth]{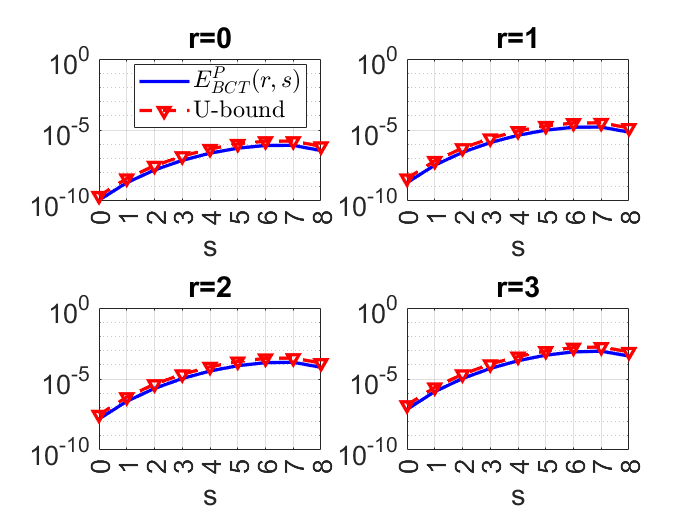}
\end{subfigure}
\begin{subfigure}[b]{\linewidth}
\includegraphics[width=\linewidth]{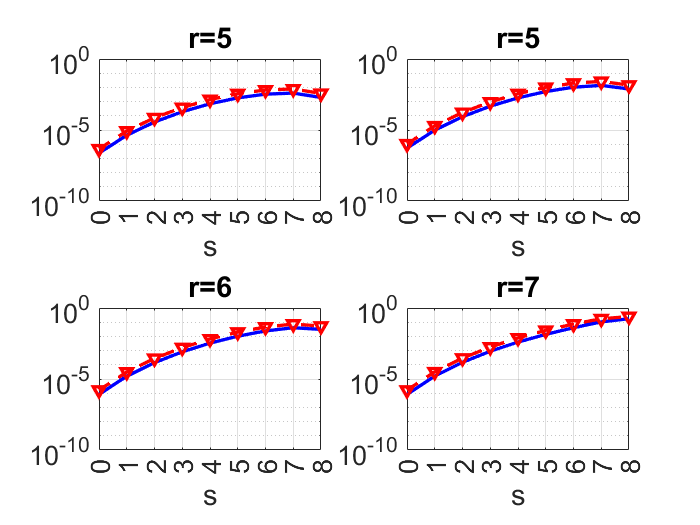}
\end{subfigure}
\begin{subfigure}[b]{0.5\linewidth}
\includegraphics[width=\linewidth]{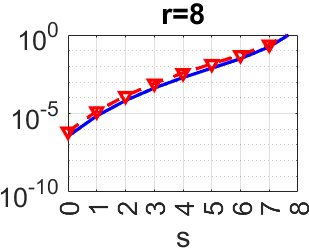}
\end{subfigure}
\caption{Plot of $\text{E}_{BCT}^P(r,s)$ in blue for the DPSS waveform. Upper bound in red.}\label{F_sig_XCORR_P}
\end{figure}
Unlike in the time and frequency domain cases, no simple analytical expressions exist for DPSSs; most treatments in the literature involve bounds on the behavior of the sequences\cite{said2023non}.
We follow the same procedure by first numerically finding  $\text{E}_{BCT}^P(r,s)$ and plotting the values in Fig. \ref{F_sig_XCORR_P}.
The trend shows that cross-correlations involving lower order DPSSs lead to lower tail energy values, and the opposite for high order DPSSs.  \textit{For the majority of sequence pairs}, the tail energy in the prolate domain reaches significantly smaller values than in time domain and frequency domain signaling.

\subsection{ISI Energy vs. Resource Utilization}

Based on the results from the previous section, the components of a given basis that leak the most for a half sample shift were identified. For both no precoding and DFT precoding, the high frequency subcarriers contribute the most to fractional leakage. For DPSS precoding, the high order DPSS components contribute the most to fractional leakage. Thus, in the tradeoff between resource utilization and ISI, we know for each basis how to minimize the leakage with the minimum sacrifice in resources.

An upper bound on the total ISI energy for a given channel can be obtained from the following lemma.

\begin{lemma} {(ISI Energy Upper Bound)} \label{lemma1_statement}
The ISI energy due to a channel path with delays $(\tau_0,..,\tau_{P-1})\in \mathbb{R}$, and corresponding path gains $(g_0,..,g_{P-1})$ and a signaling basis with cross-correlation tensor $\mathbf{C}$ is given by

\begin{equation} \label{lemma1_eq}
\text{E}^{ISI}\leq 4\sum_{r,s}\sum_{p=0}^{P-1}\sum_{l=0}^{4N_p} \left|c_{r,s}(l;N_p)\right|^2\lambda_l(1-\lambda_l)\\
\end{equation}
where $c_{r,s}(l)=\sum_{n=-2N_g}^{2N_g}\mathbf{C}_{rs}[n]s_l^{(0.25,4N_p+1)}[2n]$, $s_l^{(0.25,4N_p+1)}$ is $l$-th member sequence from the DPSS set with parameters $0.25,4N_p+1$,  $\lambda_l$ is its corresponding eigenvalue, and $N_p = N+g-\lfloor\tau\rfloor$.
\end{lemma}
\textit{Proof}. 
See Appendix I.

For unit symbol energy, the reciprocal of the above upper bound provides a lower bound on the signal-to-intersymbol interference ratio (S2I).

As an evaluation, we consider a generic delay-dispersive channel with delay spread spanning $[0,\tau_{max}=16]$ samples, and following an exponential delay profile. We consider two channels, of varying severity, as controlled by the rate of decay of the exponential delay profile: our mild channel has tap gains decaying according to $e^{-0.5 \tau_n}$ (with uniformly random phase), and our severe channel has tap gains decaying according to $e^{-0.05\tau_n}$ where $\tau_n=0,0.1,0.2,..,15$. We also consider a case where all channel taps have integer delay, i.e. $\tau_n=0,1,..,15$.

\begin{figure}
\centering
\includegraphics[width=\linewidth]{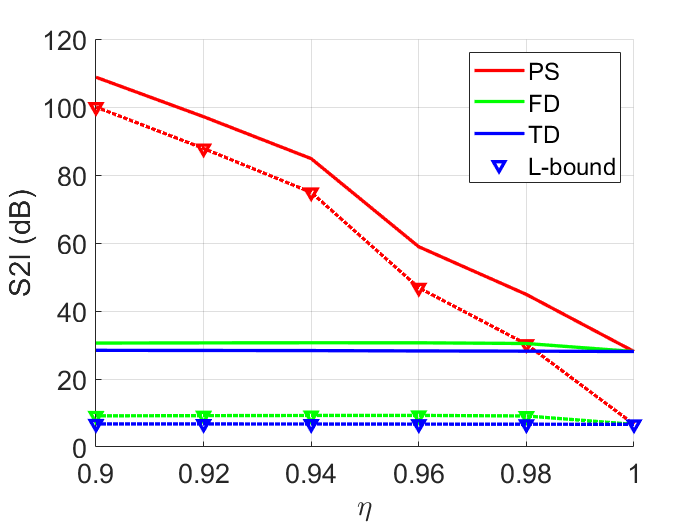}
\caption{Signal-to-inter-symbol interference versus resource utilization percentage for the \textbf{mild channel}.  Frequency domain (FD), time domain (TD) and  prolate domain (PS) signaling depicted by green, blue and red curves respectively. (a) Channel with fractional taps, (b) Channel with integer taps.}\label{S2IBI_exp_pnt5_plots}
\end{figure}

Figure \ref{S2IBI_exp_pnt5_plots} shows the variation of the signal-to-inter-symbol interference (S2I) (dB) with resource utilization $\eta=M/N$ for the mild channel case. 

In Fig.  \ref{S2IBI_exp_pnt5_plots}, at $\eta=1$ all three signaling domains have the same S2I $~28.7$ dB which can be explained by the fact that the three waveforms are \textit{complete} orthonormal bases and thus are equivalent when $\eta=1$. 

For $\eta<1$, S2I rises rapidly for the PS domain waveform at a rate of $~12$ dB per $0.02$ reduction in $\eta$ with an S2I reaching up to $110$ dB. On the other hand, DFT precoding (TD) and no precoding (FD) S2I rise at a much slower rate, with an S2I reaching up to $30$ dB for TD, and $31$ dB for FD at $\eta=0.9$. Lower bounds for S2I based on the ISI upper bound given in \eqref{lemma1_eq} are shown by black markers on top of dashed lines with the same color as the bounded S2I for a given domain. In general the bound is not very tight, however, it closely follows the general trend of the true S2I shown in solid lines. We note that ISI is effectively the sum of squares of the samples of a \textit{subsampled version} of the band-limited correlation tail sequence. This explains why the ISI upper bound is not expected to be very tight and as a consequence the S2I lower bound will also not be very tight.

\begin{figure}
\centering
\includegraphics[width=\linewidth]{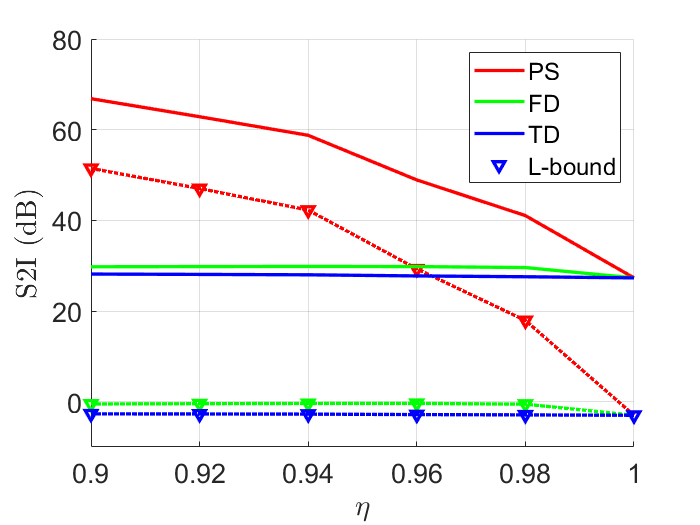}
\caption{Signal-to-inter-symbol interference versus resource utilization percentage for the \textbf{severe channel}.  Frequency domain (FD), time domain (TD) and  prolate domain (PD) signaling depicted by green, blue and red curves respectively. (a) Channel with fractional taps, (b) Channel with integer taps.}\label{S2IBI_exp_pnt05_plots}
\end{figure}


\section{Results}

\begin{figure}[h!]
\centering 
\includegraphics[width=0.8\linewidth]{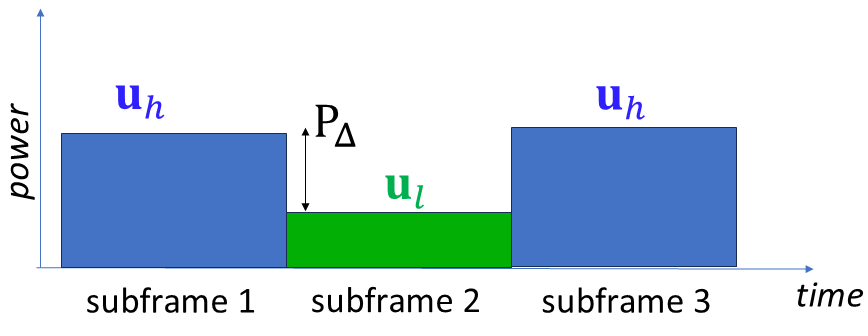}
\caption{Multi-user transmission consisting of three subframes.}\label{mu_3_subfrm}
\end{figure}
We numerically evaluate performance 
in terms of SER vs. SNR across three types of precoding over OFDM: No precoding, DFT precoding (SC-FDMA), and DPSS precoding. The simulation results are obtained by averaging over $1000$ realizations with parameters chosen according to Table \ref{table:1}.
The chosen simulation parameters are representative of a 5G NR LTE-M for M2M scenarios in outdoor environments.

\begin{table}[h!]
\centering
 \begin{tabular}{c c} 
 \hline
\multicolumn{2}{c}{Transmission Parameters}  \\  
 \hline\hline
Sampling rate & $1.92$ MHz\\
 Sub-carrier spacing & $15$ kHz\\
 FFT size & 128\\
 CP length & $\lceil 1.92 MHz \times \tau_{max}\rceil$\\
 Modulation & QPSK\\
 Time duration & 3 sub-frames (42 symbols)  \\
 User power difference $(P_{\Delta})$ & $0$dB, $10$dB\\
 Precoding & None, DFT (SC-FDMA), DPSS  \\  
 Resource utilization ($\eta$) & $1, 0.98, 0.95$ \\
 \hline
 \multicolumn{2}{c}{Channel Parameters}  \\ 
  \hline\hline
  Channel model & CDL-C\\
  Delay spread & $200$ ns, $1000$ ns\\
  \end{tabular}
 \caption{Simulation Parameters}
\label{table:1}
\end{table}

The time duration of the simulation is 3 sub-frames, each subframe consisting of 14 symbols as per the 5G standard, and each symbol comprised of $128$ samples. The first and third subframes corresponding to high-power users denoted $\mathbf{u}_h$, and the second subframe $u_l$ corresponds to a low power user as visualized in Fig. \ref{mu_3_subfrm}.

The following figures show the SER vs. SNR performance of user $\mathbf{u}_l$ as depicted in Fig. \ref{mu_3_subfrm} which is assigned subframe 2. Subframes 1 and 3 are assigned to other users, operating at a power $P_{\Delta}$ higher than $\mathbf{u}_l$. The three precoding schemes described in section III are considered. No precoding is shown by the green curves, DFT precoding by the blue curves and DPSS precoding by the red curves. The resource utilization $\eta=\lfloor\frac{M}{N}\rfloor$ is varied for each type of precoding across three values $[0.95,0.98,1]$ which are depicted by circle markers, square markers, and inverted triangle markers, respectively.

\begin{figure}[h!]
\centering 
\includegraphics[width=1\linewidth]{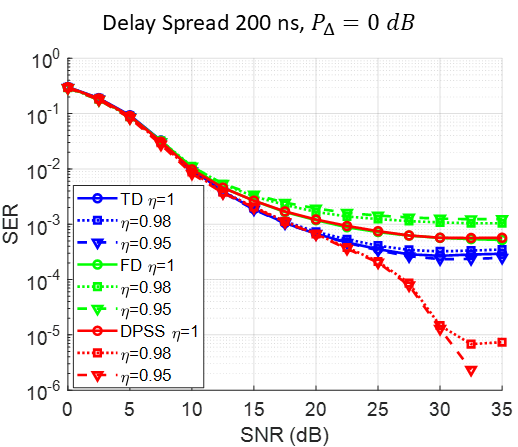}
\caption{SER v.s. SNR in channel with mild delay spread, $200$ ns for 
three types of precoding: No precoding in green, SC-FDM precoding in blue and DPSS precoding in red. Adjacent time user operating at same power level.
}\label{dsprd_200ns_P_dlt_0dB}
\end{figure}

\begin{figure}[h!]
\centering 
\includegraphics[width=1\linewidth]{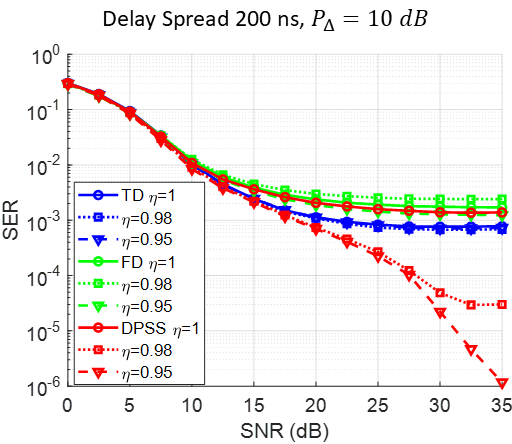}
\caption{SER v.s. SNR in channel with mild delay spread, $200$ ns for 
three types of precoding: No precoding in green, SC-FDM precoding in blue and DPSS precoding in red. Adjacent time user operating at $10$ dB power level.
}\label{dsprd_200ns_P_dlt_10dB}
\end{figure}

 \Cref{dsprd_200ns_P_dlt_0dB,dsprd_200ns_P_dlt_10dB} correspond to a channel with mild delay spread, $200$ ns. 

For the case of $P_{\Delta}=0$ dB in Fig. \ref{dsprd_200ns_P_dlt_0dB}, we can see that for SNRs up to $15$ dB there is no significant difference in performance between all three precoders and for all $\eta$ values. For $\eta=1$, full utilization, as SNR is increased beyond $15$ dB, SER starts saturating for all three precoders settling at an error floor of $2\times 10^{-4}$ for DFT precoding, $6 \times 10^{-4}$ for DPSS precoding, and no precoding. As $\eta$ is lowered, SER performance for no precoding becomes worse compared to the case of $\eta=1$ at low SNRs, and saturating at high SNRs to a higher error floor of $10^{-3}$. For DFT precoding, lower $\eta$ does not cause a significant change in performance. For DPSS, for $\eta\leq0.98$, up to $25$ dB DPSS precoding performance remains similar to DFT precoding. After $25$ dB, DPSS SER drops rapidly and saturates at $10^{-5}$ at $35$ dB for $\eta=0.98$, while for $\eta=0.95$ there is no saturation.

When the power of user $\mathbf{u}_h$ is raised by $P_{\Delta}=10$ dB, we can see in Fig. \ref{dsprd_200ns_P_dlt_10dB} the same general trends. However, now DFT precoding SER saturates at $10^{-3}$, while for DPSS precoding the SER saturation level at $\eta=0.98$ rises to $2\times 10^{-5}$, while there is no saturation for $\eta=0.95$.

The error floors experienced are due to the ISI from a user adjacent in time. However, DPSS precoding has a significantly lower error floor compared to no precoding and DFT precoding. This advantage of DPSS becomes more prominent in scenarios where adjacent users in time are operating at high power levels.

\begin{figure}[h!]
\centering 
\includegraphics[width=1\linewidth]{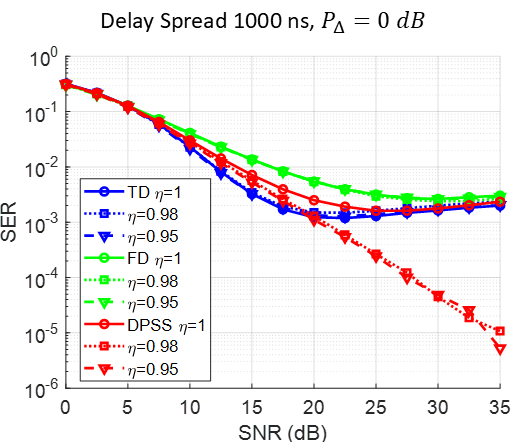}
\caption{SER v.s. SNR in channel with severe delay spread, $1000 $ ns for 
three types of precoding: No precoding in green, SC-FDM precoding in blue and DPSS precoding in red. Adjacent time user operating at same power level.
}\label{dsprd_1000ns_P_dlt_0dB}
\end{figure}

\begin{figure}[h!]
\centering 
\includegraphics[width=1\linewidth]{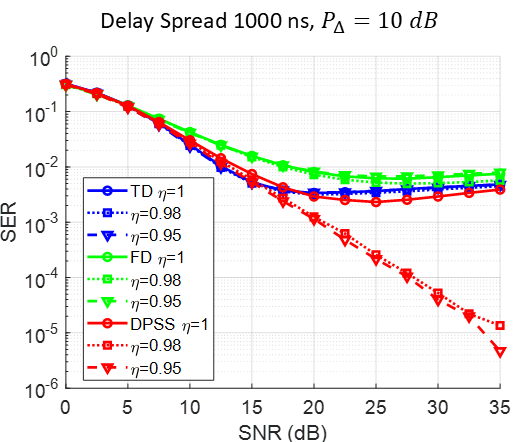}
\caption{SER v.s. SNR in channel with severe delay spread, $1000$ ns for 
three types of precoding: No precoding in green, SC-FDM precoding in blue and DPSS precoding in red. Adjacent time user operating at $10$ dB power level.
}\label{dsprd_1000ns_P_dlt_10dB}
\end{figure}

 \Cref{dsprd_1000ns_P_dlt_0dB,dsprd_1000ns_P_dlt_10dB} correspond to a channel with severe delay spread, $1000$ ns. 
For $P_{\Delta}=0 $ dB, we can see that the performance with no precoding is the worst. DFT precoding saturates at SER $10^{-3}$ at SNR $20 $ dB. While DPSS precoding performs worse than DFT precoding for $\eta=1$, as $ \eta$ is lowered the error floor is removed and SER steadily reduced by approximately an order of magnitude every $7$ dB. As the power level of the other user is increased by $10$ dB, the DFT precoding error floor rises to $3\times 10^{-3}$ while performance of DPSS precoding remains essentially the same as it was in the $P_{\Delta}=0$ case. 

\section{Conclusion}
OFDM remains the waveform of choice for 5G and beyond 5G systems. Different numerologies have been introduced in 5G NR as an enhancement to OFDM to accommodate different use cases. Service cases such as MTC involve the operation of a large number of heterogeneous users operating at different power levels and channel conditions. Due to the poor spectral confinement of OFDM, resource allocations for MTC will require large unutilized guard regions in frequency to prevent inter-carrier interference (ICI). Alternatively, sharply confined frequency allocation using sharp filters can be used. However, such filters will lead to high ISI in time due to the longer impulse responses of the filters.
Iin this work, we provide analytical expressions for quantifying the ISI when a sharp low pass filter is used, i.e., sinc pulse shaping to avoid wasteful frequency guard bands. The ISI is shown to be impacted by the choice of the orthogonal precoder used on top of OFDM, which effectively transforms the OFDM waveform basis to a different basis.  We provide an upper bound on the ISI for arbitrary orthogonal precoder choices. Finally, we demonstrate that a precoder resulting in a waveform with DPSS basis can minimize ISI for a minimum sacrifice in resources. The interference addressed in this work can be extended to other forms of interference such as intra-symbol interference  and inter-waveform interference, for example inter-carrier interference (ICI). 
We plan to address such aspects in our future works.



%
\IEEEpeerreviewmaketitle

\subsection{Appendix I} \label{Correlation_tail_bound}

Starting from the definition of $E^{ISI}(\tau_p)$ in \eqref{ISI_energy2},
\begin{equation}\label{per_path_ISI_energy}
\begin{split}
\text{E}^{ISI}(\tau_p)&=\sum_{r,s} \sum_{l'\neq 0} \left|\left\lbrace\mathcal{B}_{0.5}^{\tau_p} \mathbf{C}_{rs}\right\rbrace[l'N]  \right|^2\\
&=\sum_{r,s} \sum_{l'\neq 0} \left|\left\lbrace\mathcal{B}_{0.5}^{\Delta\tau_p} \mathbf{C}_{rs}\right\rbrace[l'N+\lfloor\tau_p\rfloor]  \right|^2\\
&\leq\sum_{r,s} \sum_{l'\notin [-(N-1),(N-1)]} \left|\left\lbrace\mathcal{B}_{0.5}^{\Delta\tau_p} \mathbf{C}_{rs}\right\rbrace[l+\lfloor\tau_p\rfloor]  \right|^2\\
&=\sum_{r,s} \sum_{l'\notin [-(N-1)-\lfloor\tau_p\rfloor,(N-1)-\lfloor\tau_p\rfloor]} \left|\left\lbrace\mathcal{B}_{0.5}^{\Delta\tau_p} \mathbf{C}_{rs}\right\rbrace[l]  \right|^2\\
&\leq\sum_{r,s} \sum_{l'\notin [-(N-1)+\lfloor\tau_p\rfloor,(N-1)-\lfloor\tau_p\rfloor]} \left|\left\lbrace\mathcal{B}_{0.5}^{0.5} \mathbf{C}_{rs}\right\rbrace[l]  \right|^2\\
&=4\sum_{r,s}\sum_{l=0}^{4N_p} \left|c_{r,s}(l;N_p)\right|^2\lambda_l(1-\lambda_l)
\end{split}
\end{equation}
where $N_p=N-1-\lfloor\tau_p\rfloor$.

The inequality in the sixth line is based on our conjecture that a half sample shift results in the highest tail energy compared to all other fractional sample shifts $\tau_p<0.5$. Finally, the last line is due to (19) from Theorem 1 in \cite{10510883}.

\subsection{Appendix II} \label{SC_FDM}
Let $z=e^{j2\pi}$
\begin{equation}\label{crss_ambg_TD}
\begin{split}
 &\mathbf{c}^T_{r,s}[p] \\
 &= \frac{1}{\eta N^2} \sum_{n=p-(N-1)/2}^{(N-1)/2}\mathbf{o}_r^*[n]\mathbf{o}_s[n-p] \\
 &= \frac{1}{\eta N^2} \sum_{n=p-\frac{N-1}{2}}^{\frac{N-1}{2}} \sum_{l'= -\frac{M-1}{2}}^{\frac{M-1}{2}} z^{-\frac{nl'}{N}}z^{\frac{l'r}{M}} \sum_{k'=-\frac{M-1}{2}}^{\frac{M-1}{2}} z^{\frac{(n-p)k'}{N}}z^{-\frac{k's}{M}}  \\
 &= \frac{1}{\eta N^2} \sum_{l',k'=-\frac{M-1}{2}}^{\frac{M-1}{2}}z^{\frac{l'r}{M}}z^{-\frac{k's}{M}}\sum_{n=p-\frac{N-1}{2}}^{\frac{N-1}{2}} z^{-\frac{nl'}{N}}  z^{j\frac{(n-p)k'}{N}}\\
 &= \frac{1}{\eta N^2} \times \\
 &\sum_{l',k'=-\frac{M-1}{2}}^{\frac{M-1}{2}}z^{\frac{l'r}{M}}z^{-\frac{k's}{M}}z^{-\frac{ (l'+k') p}{N}}\sum_{n=-\frac{N-1-p}{2}}^{\frac{N-1-p}{2}}z^{-\frac{ (l'-k') n}{N}}\\
  &= \frac{1}{\eta N^2}  \sum_{l',k'=-\frac{M-1}{2}}^{\frac{M-1}{2}}z^{\frac{rl'-k'(s+p\eta)-\eta(l'+k') p}{M}}\frac{\sin(\frac{N-p}{N}\pi (l'-k'))}{\sin(\frac{\pi(l'-k')}{N})}\\
  \end{split}
\end{equation}


%

\balance

\bibliographystyle{IEEEtran}
\bibliography{IEEEabrv,references.bib}

\end{document}